\documentclass[aip,reprint]{revtex4-1}
\usepackage[english]{babel}
\usepackage{float}
\usepackage{amsmath}
\usepackage{amssymb}
\usepackage{bm}
\usepackage{xfrac}
\usepackage{nicefrac}
\usepackage{soul}
\usepackage{color,soul}
\usepackage[dvipsnames]{xcolor}
\usepackage{graphicx}
\usepackage{array}
\usepackage{verbatimbox}
\usepackage{tikz}
\usepackage{setspace}
\usepackage[compact]{titlesec}

\newcommand*{\circled}[2][]{\tikz[baseline=(C.base)]{
    \node[inner sep=0pt] (C) {\vphantom{1g}#2};
    \node[draw, circle, inner sep=1. pt, yshift=1pt] 
        at (C.center) {\vphantom{1g}};}}
\newcolumntype{P}[1]{>{\centering\arraybackslash}p{#1}}
\newcolumntype{M}[1]{>{\centering\arraybackslash}m{#1}}
\newcommand{\etal}{\textit{et al}.}

\begin{document}

\title{Capture and Translocation of a Rod-like Molecule by a Nanopore: Orientation, Charge Distribution and Hydrodynamics}

\author{Le Qiao, Gary W. Slater}
\affiliation{Department of Physics, University of Ottawa, Ottawa, Ontario K1N 6N5, Canada}
\date{\today}

\begin{abstract}
We investigate the translocation of rods with different charge distributions using hybrid Langevin Dynamics and Lattice Boltzmann (LD-LB) simulations. Electrostatic interactions are added to the system using the $P^3M$ algorithm to model the electrohydrodynamic interactions (EHI). We first examine the free-solution electrophoretic properties of rods with various charge distributions. Our translocation simulation results suggest that the order parameter is asymmetric during the capture and escape processes despite the symmetric electric field lines, while the impacts of the charge distribution on rod orientation are more significant during the capture process. The capture/threading/escape times are under the combined effects of charge screening, rod orientation, and charge distributions. We also show that the mean capture time of a rod is shorter when it is launched near the wall because rods tend to align along the wall and hence with the local field lines. Remarkably, the \textit{orientational capture radius} we proposed previously for uniformly charged rods is still valid in the presence of EHI.
\end{abstract}

\maketitle


\section{{Introduction}}

The voltage-driven translocation of analytes through a nanopore has attracted a lot of attention due to its potential application to molecular detection in general\cite{reynaudSensingNanoporesAptamers2020,xueSolidstateNanoporeSensors2020,wenFundamentalsPotentialsSolidstate2021}, and DNA gene sequencing\cite{deamerThreeDecadesNanopore2016,hasnainReviewNanoporeSequencing2020a,gotoSolidstateNanoporesSinglemolecule2020} in particular. In short, an electric field is used to force the analyte through the nanopore (or nanochannel); during this translocation process, the analyte blocks part of the ionic current through the channel. A current meter is thus employed to detect and characterize the analytes, including large ions, nanoparticles, viruses, and charged polymers such as DNAs, RNAs, proteins, and other polyelectrolytes. A wide range of experimental, theoretical, and computational studies have been conducted to understand the mechanisms of translocation\cite{gotoSolidstateNanoporesSinglemolecule2020,yuanControllingDNATranslocation2020,palyulinPolymerTranslocationFirst2014,buyukdagliTheoreticalModelingPolymer2019}. 

While translocation is now better understood, our understanding of the physics of the capture process remains incomplete. This process is both subtle and complex; for instance, it is potentially affected by thermal diffusion, drift due to external perturbations, long-range hydrodynamic interactions, and fluid flow. Furthermore, different analytes react differently to the same conditions; for example, the electrophoretic dynamics of uniformly charged spherical particles are fairly straightforward, while for anisotropic objects such as rod-like polymers, the direction of net motion may not align with the local electric field. 

We previously investigated the capture of point-like particles with a focus on the definition of the capture radius\cite{qiaoVoltagedrivenTranslocationDefining2019}, the time dependence of the capture rate, the size of the depletion zone, and the effects of the boundary conditions\cite{qiaoEfficientKineticMonte2021}. We also examined the orientation of rod-like polymers during capture using simple theoretical arguments and a Langevin Dynamics (LD) simulation approach\cite{qiaoCaptureRodlikeMolecules2020} (thus neglecting long-ranged electrohydrodynamic interactions, EHI): this led us to introduce the concept of an orientational capture radius $R_\theta$. 

Waszkiewicz \etal \cite{waszkiewiczHydrodynamicEffectsCapture2021} further extended our work on rod orientation during the capture process by considering anisotropic diffusivity and wall hindered hydrodynamic interactions in their analytical and numerical calculations. They recovered the orientational capture radius we defined previously and concluded that rods do not follow field lines during capture due to the anisotropic diffusion\cite{qiaoCaptureRodlikeMolecules2020}. Furthermore, they showed that the trajectory of a rod towards the nanopore depends on its initial orientation and position because of the near-wall hydrodynamic interactions (these interactions were missing in our previous work). However, the electrostatic interactions between the rod and the ions in solution are still missing in their calculation; such interactions can change the dynamics, for instance when the rod is in the high field region near (or inside) the nanopore or when the rod is not uniformly charged. The main goal of the current paper is to examine these effects.

The availability of powerful GPUs and new hybrid simulation algorithms allows us to efficiently simulate molecular dynamics with EHI even for fairly large systems\cite{datarElectrokineticLatticeBoltzmann2017}. As an extension to our previous LD simulations and Waszkiewicz \etal's calculations, we now report the results of a study of the capture and translocation of a stiff rod molecule modeled using a coarse-grained raspberry-like structure coupled to the salt-containing solvent via a lattice-Boltzmann (LB) algorithm. In order to illustrate the impacts of charge screening, EHI and rod orientation, we first examine the free solution electrophoresis of rods with different charge distributions and compare the simulation results to those obtained when a sedimentation-like mechanical force is used. We then investigate the impact of EHI on the capture and translocation processes, with emphasis on the difference between small analytes and rods.

\section{Simulation details}
We simulate rod capture and translocation using a hybrid simulation approach that includes (1) a LD simulation algorithm for the motion of the rod and ions, (2) a  LB method for the fluid, and (3) a particle-particle-particle-mesh ($P^3M$) algorithm for the electrostatic interactions. The simulations were carried out using the ESPResSo package\cite{weikESPResSoExtensibleSoftware2019}. 

\subsection{The raspberry rod model}
\begin{figure}[htpb]
\begin{center}
\includegraphics[scale=1]{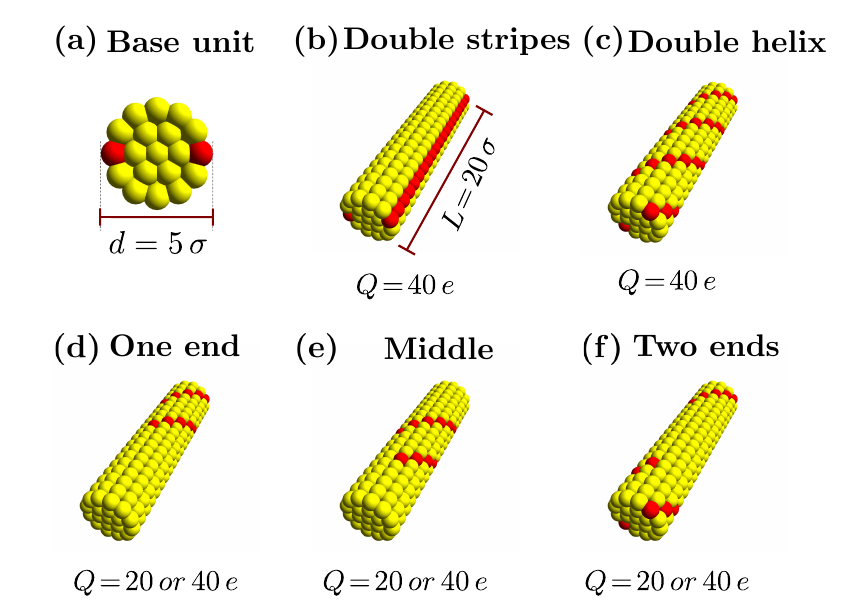}
\end{center}
\caption{Raspberry rods of length $L=20\,\sigma$ with different charge distributions (the charged beads are in red and the total charge is given by $Q$). (a) The basic disk-shaped building block. (b) The charges are lined up along two stripes. (c) A double-helix charge distribution. (d) One half of the rod is like c while the other is uncharged. (e) Similar to d, but the charged part is in the centre. (f) Same as d, but the charges are distributed only near the two ends.}
\label{Fig:dsDNA}
\end{figure}

In this section, we construct a rigid rod-like polymeric molecule of length $L \ll L_P$, where $L_P$ is the persistence length, using the "raspberry" approach\cite{ustachRaspberryModelProteinlike2016,fischerRaspberryModelHydrodynamic2015,degraafRaspberryModelHydrodynamic2015,rauDsDNAModelOptimized2017,szuttorModelingCurrentModulation2021}. The general idea of the raspberry approach is to fill the target object with enough beads (all of which interact with the LB fluid\cite{ahlrichsSimulationSinglePolymer1999}) in order to properly model its hydrodynamics properties. As shown in Fig.~\ref{Fig:dsDNA}, our rods are built by piling up $N$ raspberry disks (\ref{Fig:dsDNA}a) that each contain $1+6+12=19$ beads in three concentric layers. The beads have a radius $\frac{1}{2}\sigma$, where $\sigma$ is the fundamental length in our simulation. Therefore, the nominal diameter of the rod is $d={5\,\sigma}$ while its length is $L=N\sigma$. Although we use dimensionless units, our rod model can be used to represent a short dsDNA if we choose $\sigma=0.4\, nm$, roughly the distance between two base pairs for dsDNA. Then the diameter of the rod is $5 \times 0.4 = 2\, nm$, corresponding to the diameter of dsDNA. To make the rod rigid, all the beads are fixed in their relative position by linking them to the bead at the centre of mass via rigid bonds. 

In this paper, we also study the impact of the charge distribution on the capture process. Figures~\ref{Fig:dsDNA}~b-f show different scenarios of interest. In b-c the total charge is $Q=2N\times e$, but these charges are distributed differently along the surface of the rod. The last three rods (d-e-f) have charges over only one half of their length; if we divide these rods into four segments, the charges can symbolically be described as $[0:0:\frac{Q}{2}:\frac{Q}{2}]$, $[0:\frac{Q}{2}:\frac{Q}{2}:0]$ and $[\frac{Q}{2}:0:0:\frac{Q}{2}]$, respectively. 
In real units, if we choose $\sigma=0.4\,nm$, then the rod-like dsDNA structure has a diameter $d=2\,nm$ and the distance between basepairs is $\sigma=0.4\,nm$.

\subsection{Coupling the Lattice Boltzmann solver with LD}

We use the GPU based LB solver with D3Q19 lattice model built in the ESPResSo package to simulate the fluid, which we connect to the LD description of the rod and ions via a force coupling method\cite{ahlrichsSimulationSinglePolymer1999}. The coupling is implemented using a friction force $\bm{F}_\gamma=-\gamma (\bm{v}-\bm{u}_b)$,
where $\gamma$ is the friction coefficient, $\bm{v}$ is the bead's velocity and $\bm{u}_b$ is the fluid velocity at the bead position. An opposite force is applied to the fluid to conserve the momentum of the overall system. A zero-mean random force with a second moment that depends on temperature is added to both beads and fluid according to the fluctuation-dissipation theorem. For a bead of mass $m$, the equation of motion is thus
\begin{equation}
m \dot{\textbf{v}} = \boldsymbol{\nabla} U (\textbf{r})+ \bm{F}_\gamma + \sqrt{ 2 \gamma k_\mathrm{B} T }~ \textbf{R}(\bm{r},t),
\label{eq:Eq_lD}
\end{equation}
where $\boldsymbol{\nabla} U(\textbf{r})=\boldsymbol{\nabla}(U_\mathrm{WCA}+U_\mathrm{c}+U_\mathrm{E})$ is the sum of the conservative forces, $U_\mathrm{WCA}$ is the repulsive Weeks-Chandler-Anderson (WCA) potential between the rod beads, the ion beads and the wall:
\begin{equation}
U_\mathrm{WCA} (r) = \label{EQ:WCA}
\begin{cases}
4 \epsilon \left[ \left( \frac{\sigma}{r}\right)^{12} - \left( \frac{\sigma}{r} \right)^6 \right] +\epsilon   &\text{for } r < r_c  \\
0 &\text{for } r \geq r_c.
\end{cases}
\end{equation}
We use $\epsilon=k_BT$ as the fundamental unit of energy in our simulations, and $r_c=2^{\nicefrac{1}{6}}\,\sigma$ is the cutoff length that makes $U_{WCA}$ purely repulsive. $U_\mathrm{E} (\textbf{r})$ is the external electric potential, $U_\mathrm{c} (\textbf{r})$ is the electrostatic energy due to the charged beads. The last term is the stochastic component that models the effects of Brownian motion; the random variable $\textbf{R}(\bm{r},t)$ satisfies $\langle \bm{R}(\bm{r},t) \rangle = 0$ and
\begin{equation}
    \langle \bm{R}(\bm{r},{t}) \bm{R}(\bm{r^\prime},t^{\prime}) \rangle = \delta(t-t^\prime) \delta({\bm{r}-\bm{r^{\prime}}}) ~,
\end{equation}
where $\delta (z)$ is the Dirac delta function. The electrostatic interactions are calculated using the $P^3M$ algorithmic\cite{desernoHowMeshEwald1998a,desernoHowMeshEwald1998}. The electrostatic energy between two beads of charge $q_i$ and $q_j$ at distance $r$ is $U_\mathrm{c}(\bm{r})=c\frac{q_iq_j}{r}$, where $c=\ell_B k_BT/e^2$ with $\ell_B=e^2/4\pi\varepsilon k_BT$ the Bjerrum length and $\varepsilon$ the permittivity of the medium. The Bjerrum length is set to $\ell_B =1.8\,\sigma$ and we tune $P^3M$ to obtain an accuracy of $10^{-3}\,\epsilon/\sigma$ for the electrostatic force.

Our unit of time $\tau_o={\sigma^2\gamma}/{k_BT}$ is the time needed for a bead to diffuse over a distance $\sigma$, and the integration time step is $\Delta t=0.01\,\tau_o$ for both the LD and LB algorithms. The parameters are chosen to match the coarse grained LB dsDNA model from ref\cite{rauDsDNAModelOptimized2017,szuttorModelingCurrentModulation2021}. For instance, the friction coefficient is $\gamma=7\, m/\tau_o$, the LB kinematic viscosity is set to $\eta= 0.6\,\sigma^2/\tau_o$, the fluid density is $\rho=1\,m/\sigma^2$, and the LB lattice size is $\sigma$. The mobilities are in units of  $\mu_o=1\,\sigma^2 e/\tau_o\epsilon$. We use $C_{s}^o=0.0385\,\sigma^{-3}$ as the unit of concentration for single valence salts (this corresponds to $1 \,mol/L$ when $\sigma=0.4\,nm$). Note however that our goal is to study the impact of hydrodynamics rather than match our data to a specific experimental system.

\section{Free solution electrophoresis}
\label{sec:freeE}
In this section, we compare the electrophoresis of rods in free solution (with different charge distributions and in various salt concentrations) to the drift motion of the same rods under an equivalent mechanical force. 

In the presence of a uniform electric field, the cloud of counterions (which has a thickness $\lambda_D=\sqrt{\tfrac{\varepsilon k_B T}{2e^2C}}$, the Debye length for single valence salts; as an example, we obtain $\lambda_D\approxeq 1.75\,nm$ when $C_s=0.03\,C_{s}^o=0.03\, M$) surrounding the analyte moves in the direction opposite to the latter; the net motion of the analyte thus depends on both its size and $\lambda_D$. Note that in the case of rods, Manning condensation\cite{manningLimitingLawsCounterion1981a,manningLimitingLawsCounterion1984} may reduce the linear charge density to $e/\ell_B$ if it exceeds this critical value.

\begin{figure}[htpb]
\begin{center}
\includegraphics[scale=1]{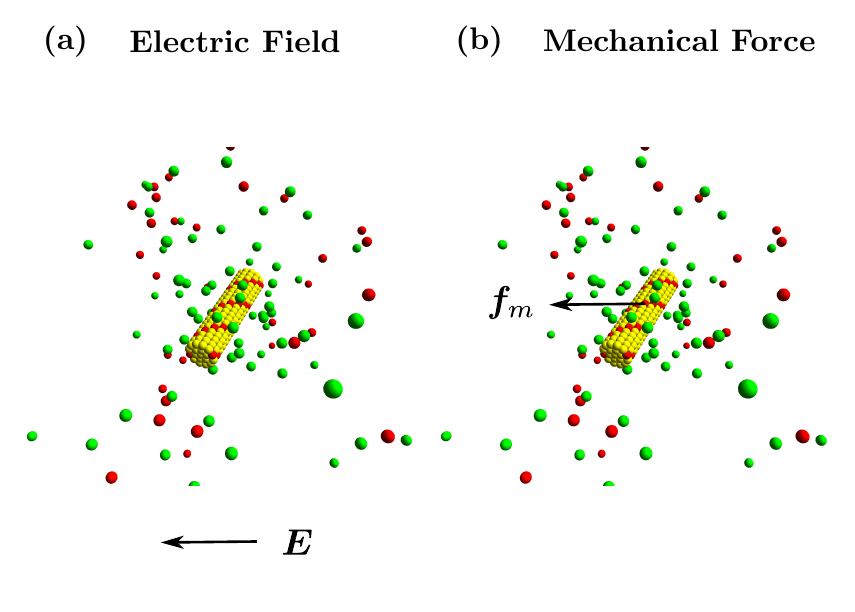}
\end{center}
\caption{Our two free-solution simulation schemes. The green and red beads are counterions and coions, respectively. (a) An electric field $\bm{E}$ affects all charged components. (b) A mechanical force is applied only to the charged beads located on the rod.}
\label{Fig:force_experiments}
\end{figure}

We carry out two different types of free-solution simulations (Fig.~\ref{Fig:force_experiments}). First, we apply an electric field $\bm{E}$ to all charged beads (including the ions in solution) to simulate electrophoresis. And then we repeat the simulations while only applying the field to the charged beads on the rod  (a weak force must also be applied to the fluid in the opposite direction in order to conserve momentum); this is equivalent to applying a mechanical force $\bm{F_m}=Q\bm{E}$ to the rod (all the other electrostatic interactions are kept). We setup the simulations with periodic boundary conditions, the box size $L_x=L_y=L_z=6L$ is large enough to remove the finite size effect due to the long range interactions cross the periodic box boundaries (data not shown).

For our purposes here, we define the electrophoretic mobility as the constant linking the mean magnitude of the instantaneous velocity and the magnitude of the applied field: 
\begin{equation}
\label{eq:mu_e}
    \langle \vert \bm{v_e}  \vert\rangle=\mu_e
    \vert\bm{E}\vert.
\end{equation}
Taking the norm ($\vert \ldots\vert$) is not necessary when the velocity and the force point in the same direction, but this is not always the case for rods when EHI effects are included, as we shall see. Similarly, we define the friction coefficient in the presence of the external mechanical force using the expression
\begin{equation}
\label{eq:mu_m}
    \vert\bm{F_m}\vert=\gamma_m \langle \vert \bm{v_m}  \vert\rangle~.
\end{equation}
To characterize rod orientation, we use the order parameter\cite{qiaoCaptureRodlikeMolecules2020} 
\begin{equation}
\label{eq:order}
      \Theta = \tfrac{1}{2} \left[ {3\langle {\cos^2 \theta}\rangle-1} \right]
\end{equation}
where $\theta$ is the angle between the direction of the force and the rod's principal axis.

\begin{figure}[htpb]
\begin{center}
\includegraphics[scale=1]{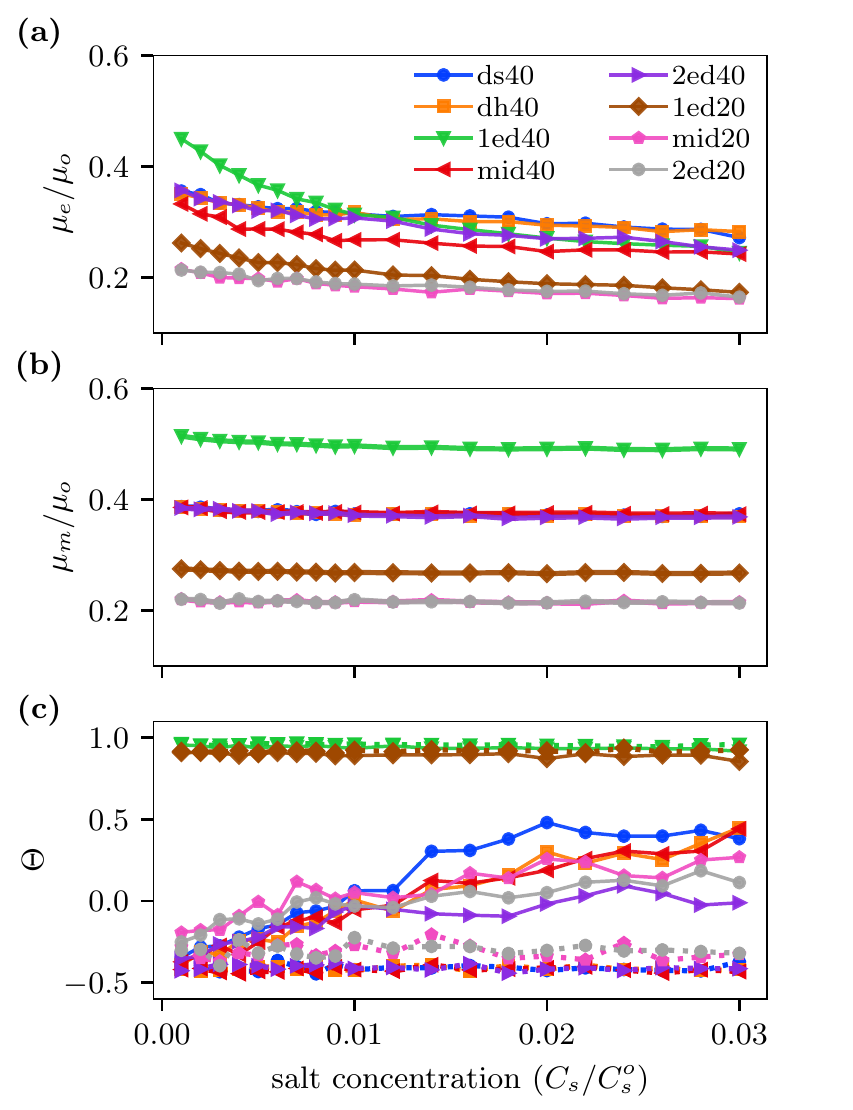}
\end{center}
\caption{(a) Scaled electrophoretic mobility $\mu_e/\mu_o$; (b) scaled mechanical mobility $\mu_m/\mu_o$; and (c) order parameter $\Theta$, \textit{vs} the salt concentration $C_s/C_{s}^o$, for different charge distributions. The results are averages over long trajectories ($10^5$ integration steps) starting with 10 random initial orientations. In panel (c), the dashed (solid) lines correspond to the mechanical (electrical) case. Legend: The codes read AB, where the A is for charge location (ds for double-striped, dh for double helix; 1ed and 2ed for one or two end; mid for middle) and B is the total bare charge $Q/e$. The electric/mechanical force applied to the charged sites on the rod is of magnitude $1 \,\frac{\epsilon}{\sigma e}$. }
\label{Fig:mu_C}
\end{figure}

Figures~\ref{Fig:mu_C}a and b show that the rods' electrophoretic mobility $\mu_e$ and equivalent mechanical mobility $\mu_m=Q/\gamma_m$ behave differently when we change the salt concentration $C_s$: while $\mu_e$ decreases when we increase $C_s$, as expected, $\mu_m$ is unaffected.  Figure~\ref{Fig:mu_C}c, together with Figs.~\ref{Fig:mu_C}a-b and Table \ref{tab:1} (which gives the mobilities and orientations for $C_s=0.03\,C_{s}^o$), clarify the physics of the problem. Let us summarize the main elements:

\begin{itemize}
    \item When the force is applied at only one of its ends, the rod tends to align along the direction of the field ($\Theta \to 1$).
    
    \item Since mechanical friction is smaller when the rod is aligned\cite{tiradoComparisonTheoriesTranslational1984} (i.e., $\gamma_\parallel<\gamma_\perp$), $\mu_m$ increases with orientation. For example, $\mu_m$ is $\approx 1.3$ times larger when only one end is pulled by the mechanical force -- Table \ref{tab:1}.
    
    \item Despite the increased orientation obtained when the electric charge is at only one end, $\mu_e$ is actually smaller than for a uniformly charged rod. This is due to the fact that Manning condensation is more intense for partially charged rods (1ed40, 2ed40, mid40) since they have a higher local linear charge density.
    
    \item When considering only partially charged rods, the electrophoretic mobilities are ranked in this order: $\mu_e({1ed})>\mu_e({2ed})>\mu_e({mid})$. This indicates that the charge condensation and the electrophoretic force have different weights along the rod\cite{mccormickMolecularEndEffect2005,chubynskyTheoryEndlabeledFreesolution2014a}. Given the same charge distribution, the mobility is expected to be essentially the same when the charge is reduced by half because of Manning condensation. In fact, the mobility is reduced by about $\frac{1}{3}$ in our simulations; this is due to the small differences in orientation after charge reduction (see Table \ref{tab:1}) and finite-size effects.
    
    \item When the mechanical force is applied uniformly along the rod, the latter tends to orient perpendicular to the force ($\Theta \to -\frac{1}{2}$), similar to what was reported in\cite{shinStructureDynamicsDilute2009,hamidNumericalStudySedimentation2020}.
    
    \item When we electrophorese a uniformly charged rod, on the other hand, we observe an interesting transition: the rod orients perpendicular to the field at low salt concentration (in which case $\lambda_D$ is too large to be a factor) while $\Theta_e \approx 0-0.5$ at high salt concentration. This is a good example of the Smoluchowski-H\"uckel transition\cite{soumpasisDebyeHuckelTheory1978,morrisonElectrophoresisParticleArbitrary1970} that we expect when the salt concentration is changed.
\end{itemize} 

\begin{table}[htbp]
\small
  \caption{\label{tab:1} Scaled mobilities $\mu_e/\mu_o\, (\pm0.003)$ and $\mu_m/\mu_o\, (\pm0.001)$, and order parameters $\Theta_e\,(\pm0.05)$ and $\Theta_m \,(\pm 0.01)$, for different charge distributions in a salt concentration $C_s=0.03\,C_{s}^o$.}
  \begin{tabular*}{0.48\textwidth}{@{\extracolsep{\fill}}llllll}
\hline
 Charge Distribution & Q/e & $\mu_e/\mu_o$ & $\mu_m/\mu_o$ & $\Theta_e$ & $\Theta_m$\\
 \hline
 Double stripes   & 40 &  0.280  &  0.374   &   0.47   & -0.37\\
 Double helix    & 40 &  0.283  &  0.369   &   0.46   & -0.42\\
 One end         & 40 & 0.246  &  0.491   &   0.92   &  0.96\\
 Middle          & 40 & 0.236  &  0.374   &   0.33   & -0.43\\
 Two ends        & 40 & 0.244  &  0.369   &  -0.05   & -0.41\\
 One end     & 20 & 0.174  &  0.268   &   0.86   &  0.93\\
 Middle      & 20 & 0.160  &  0.215   &   0.20   & -0.33\\
 Two ends   & 20 & 0.162  &  0.214   &   0.08   & -0.32\\
 \hline
  \end{tabular*}
\end{table}

\begin{figure}[htbp]
\begin{center}
\includegraphics[scale=1]{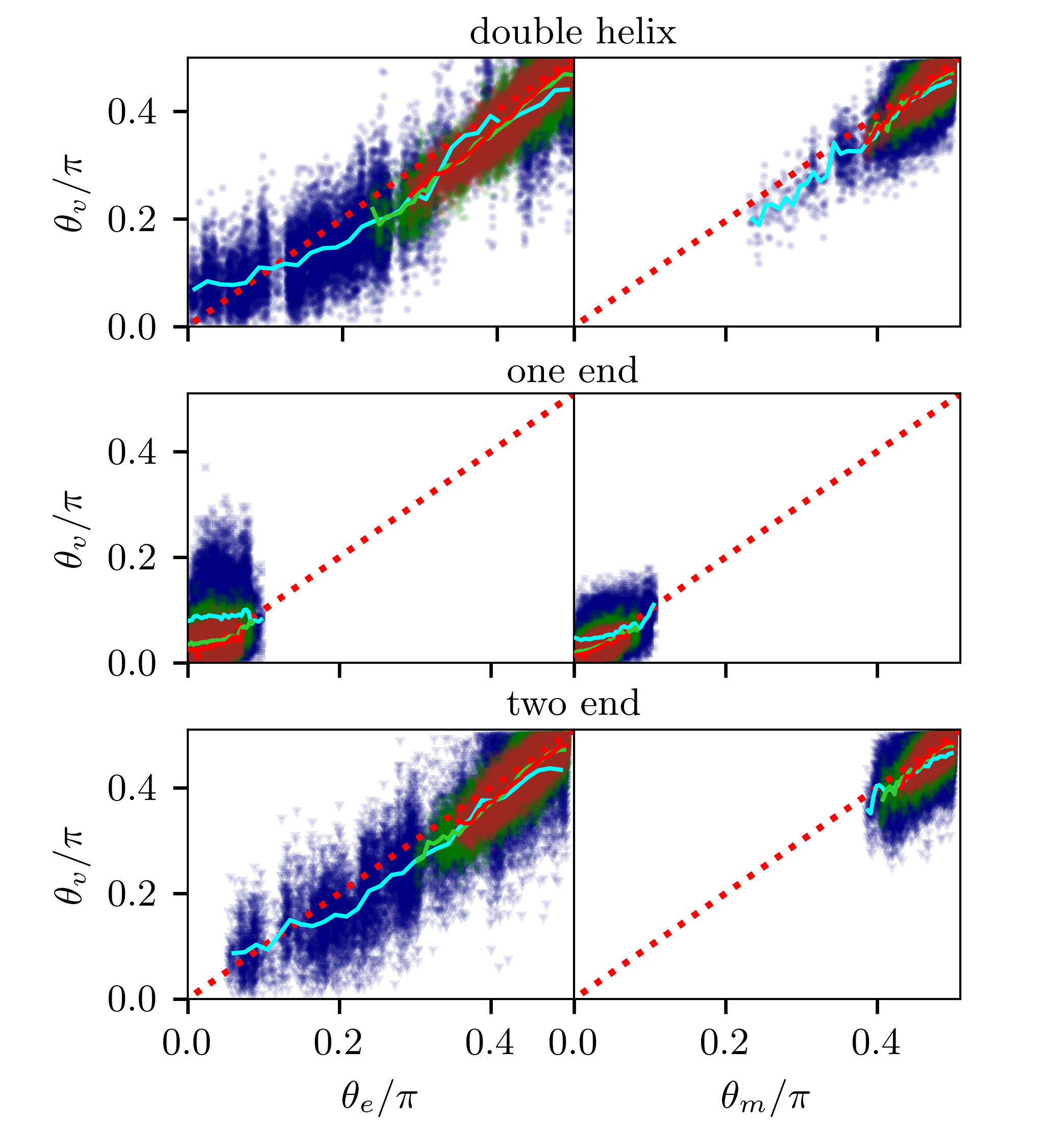}
\end{center}
\caption{Correlation between the directions of the local velocity and of the local external field/force for a rod of charge $Q=40\,e$ with different charge distributions. $\theta_v$ is the angle between velocity and the rod axis, $\theta_e$ (or $\theta_m$) is the angle between the rod axis and the direction of the electric (mechanical) force. Scattered data points are instantaneous values and the solid curves are averages. The red dashed lines show $\theta_v=\theta_e$ and $\theta_v=\theta_m$. The colors code for the magnitude of the force applied to the charged sites on the rod: the values are $\textcolor{blue}{1~ (blue)},\textcolor{green}{2~ (green)},\textcolor{red}{3~ (red)}\, \frac{\epsilon}{\sigma e}$. The salt concentration $C_s/C_{s}^o=0.03$ is used for these simulations.}
\label{Fig:correlation_v_f}
\end{figure}

We also investigated the correlation between the directions of the instantaneous velocity and of the rod axis. Figure~\ref{Fig:correlation_v_f} shows the angle between the velocity and the rod axis \textit{vs} the angle between the force and the rod axis for both electric and mechanical forces. The diagonal dashed lines correspond to the velocity being aligned with the local field. Obviously, the velocity is not perfectly parallel to the field direction, except when the rod is oriented parallel or perpendicular to the field ($\theta=0$ or $\pi/2$). 

Using the double helix (dh) charged rod as an example, the deviation reaches a maximum value of $\approx 10^{\circ}$ when the rod makes an angle of $\theta \approx \pi/3$ with the field. Moreover, when we increase the field intensity, the rod is more frequently oriented perpendicular with the field direction ($\theta \to \pi/2$) while the fluctuations due to thermal motion are much reduced; this is also observed when we use a mechanical force. 

The situation is entirely opposite for the one-end (1ed) charged rod, which tends to orient with the field/force even when the force is small. The two-end (2ed) charged rod is more or less the same as the double-helix rod, although this rod seems to prefer perpendicular orientations under a mechanical force.

Our free solution electrophoresis simulations thus indicate that a rod can orient with the field even under an uniform field at high salt concentrations. The hydrodynamic interactions can make the rod move in a direction different from that of the applied force. This last point suggests that the rod will move between field lines when the latter are converging (this is the case during the capture process), a phenomenon that we will observe in the next section.
 
\section{Orientation duration capture}
\subsection{Simulation setup}
\begin{figure}[htpb]
\begin{center}
\includegraphics[scale=1]{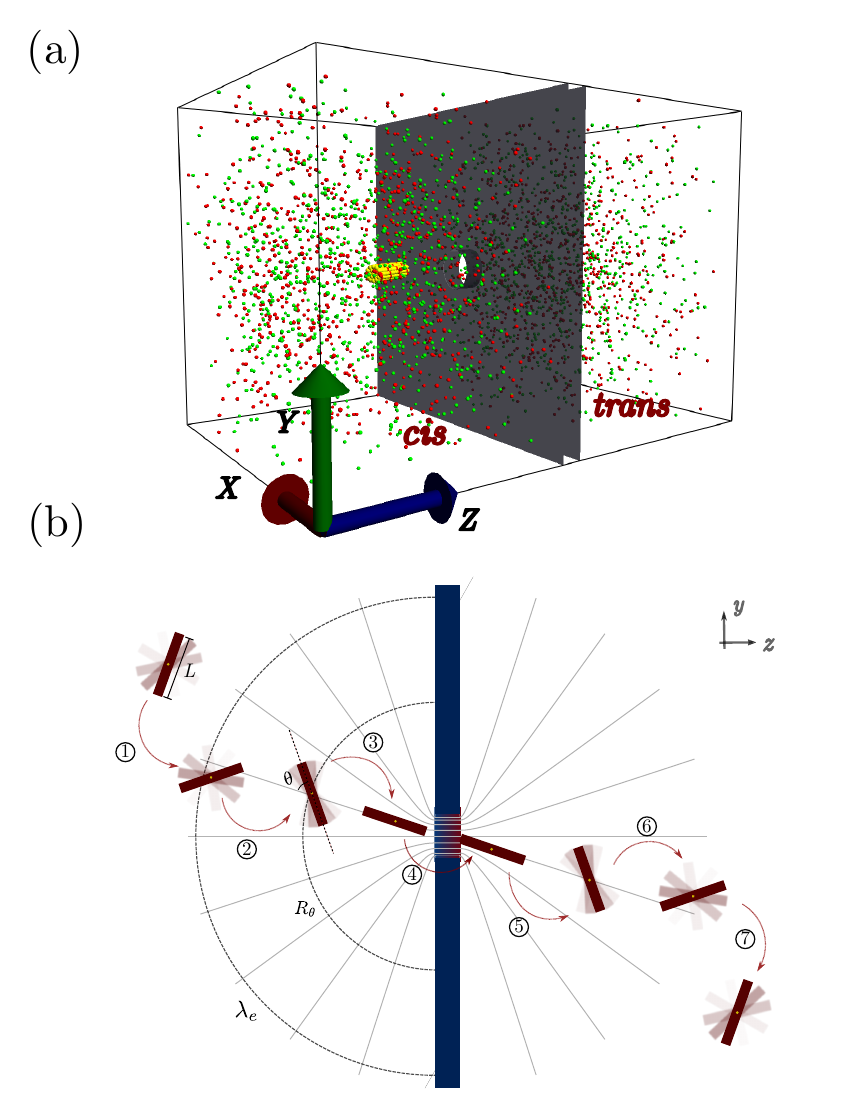}
\end{center}
\caption{(a) Snapshot of the simulation system for a double-helix rod molecule and a salt concentration $C_s=0.03\,C_s^o$. The colored beads in solution represent 
cations (\textcolor{green}{green}) and anions (\textcolor{red}{red}).~ (b) The different stages of the translocation process: \circled{1} - \circled{3} show the three phases of rod capture (diffusion $\rightarrow$ drift $\rightarrow$ drift and orientation); \circled{4} is the threading stage; \circled{5} - \circled{7} show the three phases of rod escape (drift and disorientation $\rightarrow$ drift $\rightarrow$  diffusion}).
\label{Fig:translocation}
\end{figure}

We now set up a translocation simulation system with a periodic box of size $L_x=L_y=\frac{1}{2}\,L_z=6\,L$ and an impenetrable wall with a nanopore in its centre. The radius of the pore is $r_p \! = \! \frac{5}{4}\, d= \frac{25}{4} \,\sigma$ and the length $\ell_p \! = r_p$. The charged rod is initialized on the \textit{cis} side as shown in Fig.~\ref{Fig:translocation}a. We also randomly add $N_c=2\,N$ explicit counterions beads and $N_s=2\,CV_{box}$ single valence salt ions to the system, where $C_s=0.03\,C_{s}^o$ is the salt concentration and $V_{box}=L_xL_y(L_z-\ell_p)+\pi r_p^2\ell_p$ is the accessible volume (this excludes the impenetrable wall).

The electrostatic potential outside the pore is given by \cite{farahpourChainDeformationTranslocation2013}.
\begin{equation}
 \label{seq:sol}
  V(\zeta,\beta,\phi) = \Delta V~ \frac{r_e }{r_p}~\arctan \left[ \sinh(\zeta) \right],
\end{equation}
where $\Delta V$ is the total potential difference across the device, $r_e=r_p/(\frac{2\ell_p}{r_p}+\pi)=r_p/(2+\pi)$ is the electrostatic length of the nanochannel, and  $\zeta \! \in \! (- \infty, + \infty)$, $\beta \! \in \! [ 0, \pi ]$ and $\phi \! \in \! [ 0, 2 \pi ]$ are the oblate spherical coordinates\cite{farahpourChainDeformationTranslocation2013}. The potential drop across the channel is
\begin{equation}
    \delta V =\Delta V \times \frac{2\ell_p r_e}{r_p^2},
\end{equation}
corresponding to a uniform electric field 
\begin{equation}
    E_p \! = \! \frac{\delta V}{\ell_p} \! = \! \Delta V \times \frac{2r_e}{r_p^2}.
\end{equation}

\subsection{Orientation during capture and escape}
\begin{figure}[htbp]
\begin{center}
\includegraphics[scale=1]{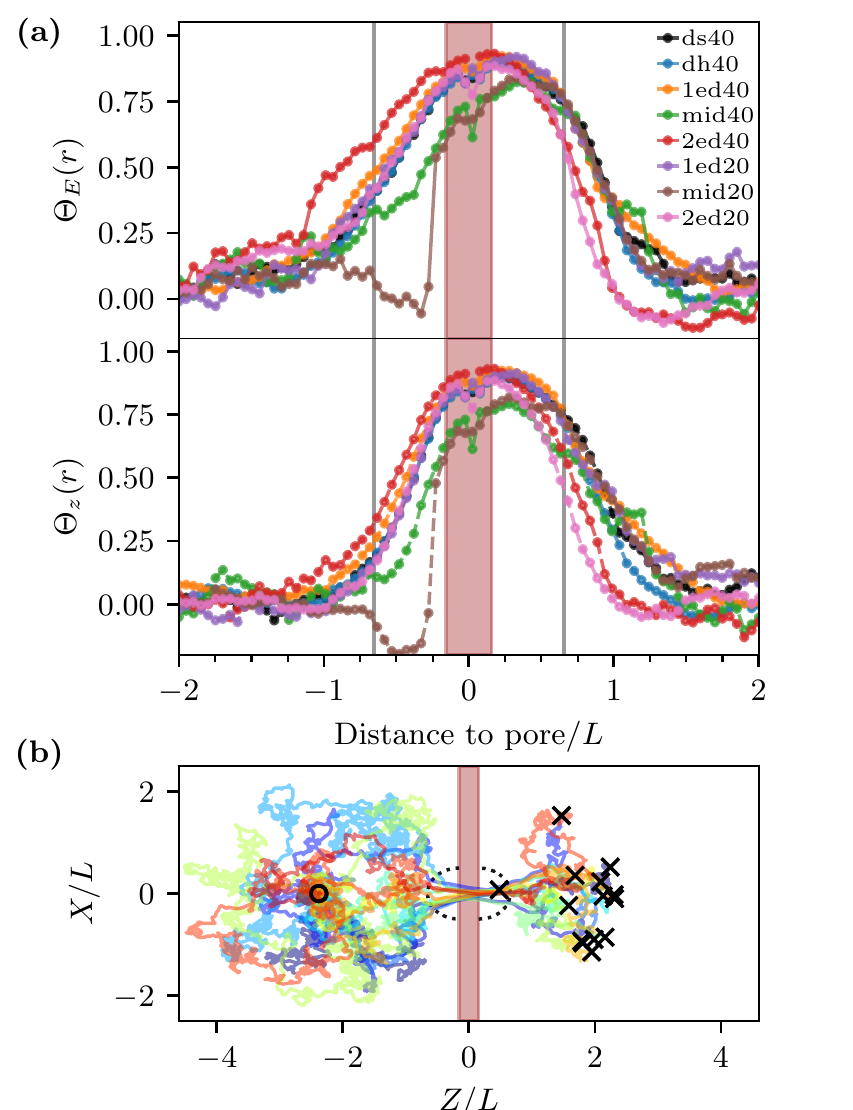}
\end{center}
\caption{(a) Order parameters $\Theta_E(r)$ and $\Theta_E(r)$  \textit{vs} scaled distance to the pore, $r/L$, for rods with various charge distributions. The rods start with random initial orientations, right above the nanopore at a distance of $r=2.2\,L$ from the centre of the pore entrance (see the {\large{$\circ$}} symbol in b below) and stop when they arrive the same distance from the centre of the pore exit on the \textit{trans} side ($\bm{\times}$). The applied potential is $\Delta V=15.6\, k_BT/e$, which corresponds to $400~mV$ at room temperature. The vertical lines are at a distance $L/2$ from the wall (the latter is marked by the shaded area). Each curve is averaged from 50 simulations using a salt concentration $C_s/C_{s}^o=0.03$. (b) Trajectories of ds40 rods projected on the $XZ$ plane. Two doted semi-cycles indicates the CM position when the rod first enter/exit the nanopore.}
\label{Fig:order}
\end{figure}

In order to illustrate the impacts of the charge distribution on the orientation of the rod during the capture, translocation and escape processes (see Fig.~\ref{Fig:translocation}b), we show how the order parameter $\Theta(r)$ depends on the radial distance to the pore ($r$) in Fig.~\ref{Fig:order}a. We present two different order parameters: $\Theta_E$ uses the angle between the rod axis and direction $\bm{\hat{E}_{CM}}$ of the field at the centre-of-mass of the rod, while $\Theta_z$ uses the pore axis $\bm{\hat{z}}$ instead. Several trajectories are shown in Fig.~\ref{Fig:order}b for the case of a double-striped rod.

Despite the fact that all of the rods are launched from the same position, right above the nanopore (the black dot in Fig.~\ref{Fig:order}b), they follow different trajectories and spread widely before arriving at the nanopore. Figure~\ref{Fig:order}a shows that the rod is oriented along the local field direction $\bm{\hat{E}_{CM}}$ and not along the nanochannel axis $\bm{\hat{z}}$. The two order parameters do not merge until the rod is engaged in the nanopore, which suggests that the rod tends to enter the pore sideways even when starting right above the nanopore. There are three reasons for this: (1) Rods diffuse in random directions before entering the high field region. (2) When a rod is already in the high-field region but not fully aligned with the local field, it tends to jump between field lines. (3) The field is higher near the pore edges\cite{qiaoVoltagedrivenTranslocationDefining2019}. The trajectories in Fig.~\ref{Fig:order}b also show that rods tend to move along the wall (note the depletion region right above the pore). The two order parameters converge at the pore because the entry process aligns the rod with the pore axis $\bm{\hat{z}}$. 

Although the field lines are identical on both sides of the wall, we observe a clear asymmetry in the rod orientation: the orientation is kept for a larger radial distance when escaping from the pore. More strikingly, there is little difference between the $\Theta_E$ and $\Theta_z$.

Rods with different charge distributions follow different $\Theta(r)$ \textit{vs} $r$ curves during the capture process. For $Q=40\,e$, the rod with charges at both ends tends to be more oriented than the rod charged at only one end, followed by the uniformly charged rods (both dh40 and ds40).  The rod with middle charges, on the other hand, is the least oriented.

When reducing the charge to $Q=20\,e$, similar results are found, except for the rod charged in the middle, which shows no orientation until it is well inside the nanopore. In the latter case, the negative values of $\Theta_z(r)$ near the nanopore pore indicate that these rods arrive misoriented and thus require large amount of time to enter the nanopore (see Table~\ref{tab:2}). Overall, the fact $\Theta_z(r)<\Theta_E(r)$ suggests that rods enter the nanopore from the side despite being launched right above it. During the escape process, the rods follow roughly the same $\Theta(r)$ curves except for the two-end charged rods, which disorient faster.

As shown in Table~\ref{tab:2}, rods charged at only one end enter the nanopore via this end while there is no preference for the other rods. The data also show that the capture time $\tau_c$ for one end of the rod to enter the nanopore from its initial position is roughly the same for rods that have the same type of charge distribution (one end and two end) but different charges ($Q=20\,e$ \textit{vs} $40\,e$), a consequence of Manning condensation. For rods that are only charged in the middle, reducing the charge from $Q=40\,e$ to $20\,e$ leads to a reduction in orientation as shown in Fig. \ref{Fig:order}; as a consequence, the rod spends more time to place one of its ends in the nanopore to complete the capture process. The same effect also explains why chains with the same total charge $Q=40\,e$ but different distributions have different capture times (\textit{e.g.}, double-helix \textit{vs} one end).

Clearly, the translocation times $\tau_t$ for rods to thread the nanopore are directly impacted by both the bare charge density and the location of this charge (note that because the wall thickness is $< L/2$, we have cases where the rod segment inside the channel is neutral during part of the translocation process). Moreover, the nature of the pore-rod and hydrodynamic interactions inside the channel may also impact the translocation times.

Finally, we see in Table~\ref{tab:2} that although the escape times $\tau_e$ for rods to move away from the pore exit to the same distance as the initial position from the pore entrance are about four times smaller than the capture times, the relative escape times are very similar except for the rod with charge on one end.

\begin{table}[htbp]
\small
  \caption{\ Probability $P$ for the rod to enter the nanopore via a pre-determined end (the charged one for the one-end cases); capture time $\tau_c$; translocation time $\tau_t$; and escape time $\tau_e$ for different rod types and nominal charges $Q$. The times are normalized by the values found for the two-stripe case (first line), i.e., $3.1(5)\times 10^4$, $2.5(2)\times 10^2$, and $0.7(3)\times 10^4\,\tau_o$, respectively.}
  \label{tab:2}
  \begin{tabular*}{0.48\textwidth}{@{\extracolsep{\fill}}llllll}
 \hline
 Charge distribution & $Q/e$ & $P$ & $\tau_c $& $\tau_t$ &$\tau_e $\\
 \hline
 Double stripes    & 40 &  1/2   &   1       &  1     & 1   \\
 Double helix   & 40 &  1/2   &  1.0      &1.0    &1.0   \\
 One end        & 40 &  1     &  1.3      &2.2    &1.1   \\
 Middle         & 40 &  1/2   &  1.4      &1.6    &1.2   \\
 Two ends       & 40 &  1/2   &  1.0      &1.1    &1.0   \\
 One end        & 20 &  1     &  1.2      &2.7    &1.3   \\
 Middle         & 20 &  1/2   &  6.0      &2.9    &1.1   \\
 Two ends       & 20 &  1/2   &  1.3      &1.5    &1.1   \\
 \hline
  \end{tabular*}
\end{table}
\subsection{Initial orientations}
\begin{figure}[htbp]
\begin{center}
\includegraphics[scale=1]{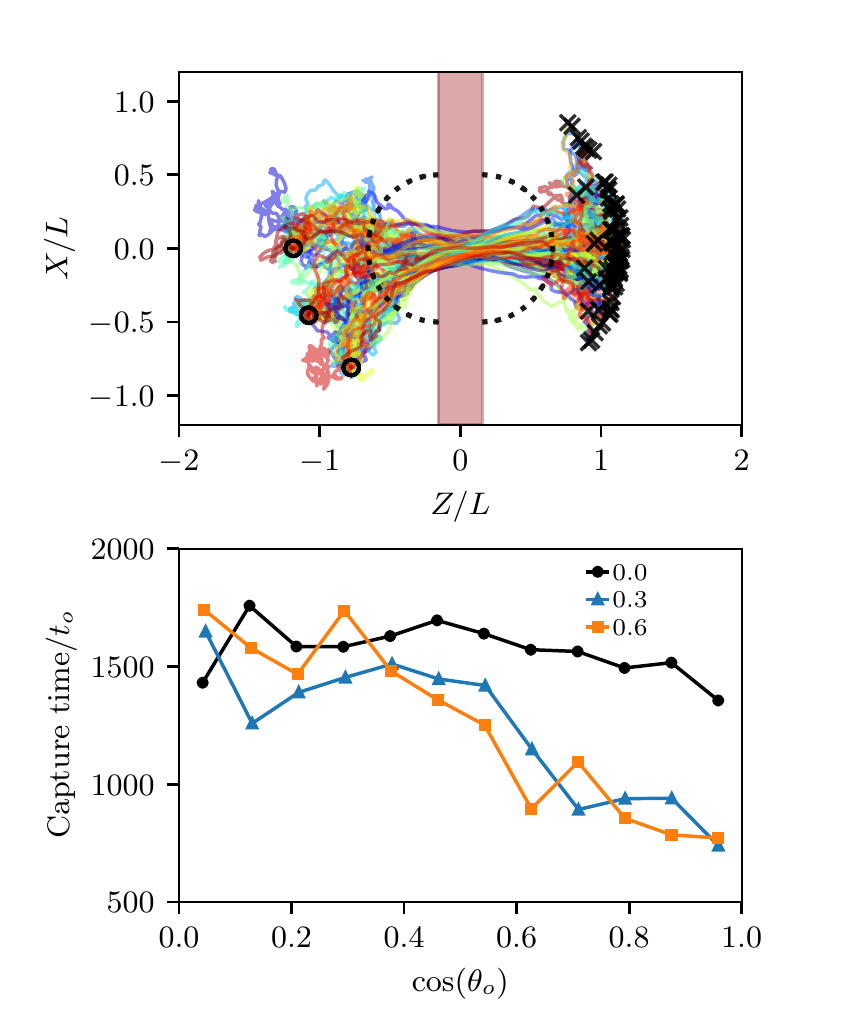}
\end{center}
\caption{(a) Centre-of-mass trajectories of double-helix rods with $Q=40\,e$ (or dh40) during translocation, projected on the $XZ$ plane. The three black empty circles ($\circ$) indicate the initial position, all at a distance $r=L$ from the pore centre at $[0,0,-\frac{1}{2} \ell_p]$ but with varying polar angles $\phi_o\in [0, 0.3 ,0.6] \times\frac{\pi}{2}$ from the pore axis. The final positions are marked with cross-marks ($\times$). Two doted half circles indicate the CM position when the rod first enter/exit the nanopore. (b) Capture time \textit{vs} initial rod orientation for the same three different initial polar angles $\phi_o$. The rod is parallel to the local field when $\cos(\theta_o)=1$ and perpendicular when $\cos(\theta_o)=0$. Each curve is an average over 150 translocation simulations; the salt concentration is $C_s/C_{s}^o=0.03$.}
\label{Fig:correlation}
\end{figure}

In this section, we examine whether the initial orientation of a rod has an impact on its capture time. We place the randomly oriented rods at a distance $L$ from the entrance of the nanopore; as shown in Fig.~\ref{Fig:correlation}a, we start them from three different angular positions (polar angles). Since they start their journey very close to the pore, we know the field gradient will modify their initial orientation well before they reach the pore.

For the rods starting right above the nanopore (black line) in Fig.~\ref{Fig:correlation}, the capture time is almost flat for all initial orientations $\theta_o$.  However, for the other two polar angles $\phi_o$, the capture time is a strong function of the initial orientation, with perpendicular orientations taking twice as much time as parallel ones. Rods that start nearly aligned with the local field direction encounter less friction from the start. These results are consistent with our previous investigations\cite{qiaoVoltagedrivenTranslocationDefining2019}. 

The capture time for the different initial angular positions converges to roughly the same value when the rod is initially perpendicular to the local field because the time for the rod to rotate and align with the field then dominates the capture time.  However, when the rods are already aligned with the field lines and start close to the wall, the presence of the wall helps the rod maintain its alignment and the capture time is shorter, similar to what is reported in ref\cite{waszkiewiczHydrodynamicEffectsCapture2021}. 

\subsection{Orientational capture radius}
We previously proposed an orientational radius $R_\theta$ to characterize the rod orientation during capture but we did not consider electrohydrodynamic interactions\cite{qiaoCaptureRodlikeMolecules2020}: this orientational capture radius depends on the field intensity and the length of the rod, 
\begin{equation}
\label{eq:orientation_radius}
    R_\theta = \left(\tfrac{1}{60}\lambda_eL^2\right)^{1/3},
\end{equation}
where $\lambda_e$ is the capture radius\cite{qiaoVoltagedrivenTranslocationDefining2019},
\begin{equation}
\label{eq:lambda_definition}
    \lambda_e =\frac{\tilde{Q} \Delta V }{k_BT}~ r_e
\end{equation}
with $\tilde{Q}$ the effective electrophoretic charge of the analyte. 

We now revisit the problem by considering the effects EHI might have on a double-helix rod. If we assume that the charge of the rod stays constant for different field intensities, the orientational radius should only depend on the applied voltage, with $R_\theta \sim \lambda_e^{\nicefrac{1}{3}} \sim \Delta V^{\nicefrac{1}{3}}$. In order to test this voltage dependence, we simulated the capture of the rod by applying different voltages $\Delta V=15.6,~ 31.2$ and $46.8\, \frac{k_BT}{e}$. Our data do indeed collapse when $\Theta(r,\Delta V)$ is plotted \textit{vs} $r/\Delta V^{1/3}$ (data not shown) despite the presence of EHI. 

A more complete investigation of the effects of the EHI on rod orientation would require that we estimate the orientation radius $R_\theta= \left(\tfrac{1}{60}\lambda_eL^2\right)^{\nicefrac{1}{3}}$. To do so, we must determine the rod's effective charge $\tilde{Q}$ in order to obtain $\lambda_e=\frac{\tilde{Q} \Delta V }{k_BT}\, r_e$. The effective charge of a spherical analyte can be estimated from its electrophoretic mobility $\mu$ using the expression $\tilde{Q}=\mu\frac{k_BT}{D}$, where $D$ is the diffusion coefficient of the analyte. The local electrophoretic mobility of the rod is given by the value obtained in the uniform electric field case in Sec. 3 and Table~1: $\mu_e\approxeq0.283\,\mu_o$. To simplify, we then assume that the rod gains full orientation (along the direction of the local electric field) immediately when it reaches the orientational radius at $r=R_\theta$. Its friction coefficient $\gamma_m=\frac{k_BT}{D}$ at this location should be obtained under a mechanical force when the rod has the mean orientation $\Theta\approxeq 1$; we thus use the friction coefficient measured when the mechanical force is applied only at one end in free solution simulations because this is the case with the largest degree of orientation. Rewriting eq.~\ref{eq:mu_m}, one gets $\gamma_m=\frac{\vert\bm{F_m}\vert}{\vert\langle\bm{v_m}\rangle\vert}=\frac{1}{\mu_m}$; therefore, the effective charge is simply given by the ratio of the two mobilities when the field and force are equal. Here this gives $\tilde{Q}/Q\approx{\mu_e(dh40)}/{\mu_m(1ed40)}\approx0.58$, or $\tilde{Q}\,=23\,e$, where $\mu_m(1ed40)\approxeq0.491\,\mu_o$. In simulations, we chose $\Delta V=15.6,~ 31.2$ and $46.8\, \frac{k_BT}{e}$, which gives capture radii $\lambda_e\approx 435.1, ~870.3$ and $1305.4\,\sigma$, respectively. Given the rod length $L=20\,\sigma$, these values correspond to $R_\theta=14.3,~ 18.0$ and $20.6\, \sigma$.

\begin{figure}[htpb]
\begin{center}
\includegraphics[scale=1]{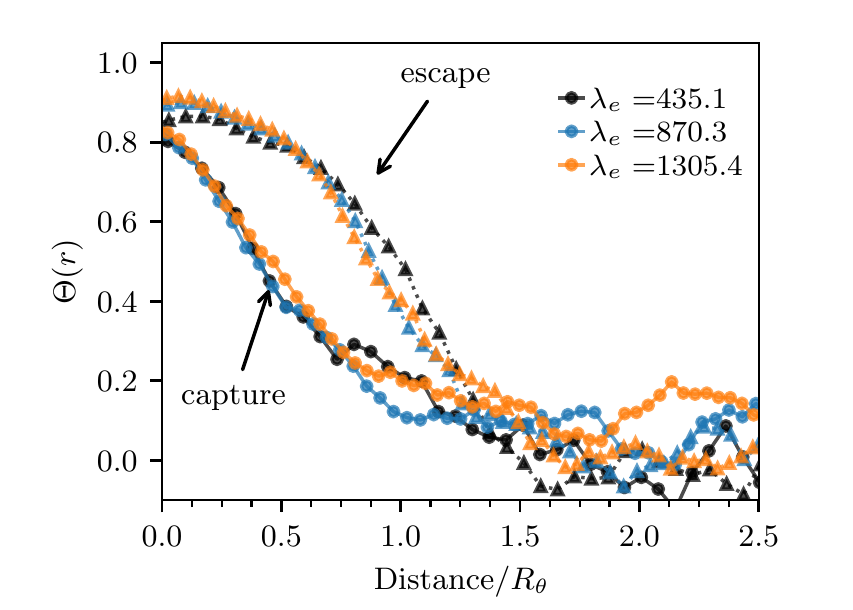}
\end{center}
\caption{Order parameter $\Theta(r)$ \textit{vs} scaled radial distance $r/R_\theta$ to the pore entrance/escape during the capture/escape processes for a double-helix rod ($Q=40\,e$) with different field intensities $\lambda_e$. We have $R_\theta=14.3,~ 18.0$ and $20.6\, \sigma$ for the cases shown here. Each curve is averaged over 50 translocation simulations using a salt concentration $C_s/C_{s}^o=0.03$.}
\label{Fig:order_field}
\end{figure}

As shown in Fig.~\ref{Fig:order_field}, the order parameters obtained at different field intensities collapse on a single curve after rescaling the distance to the pore by these estimates of $R_\theta$. Interestingly, the curve is not the same for the capture by, and exit from, the pore, again showing the asymmetry between these two processes. Nevertheless, we see that the order parameter essentially vanishes for $r > R_\Theta$; we thus conclude that despite the presence of EHI, the orientational capture radius defined previously remains valid, including for the escape process (we did not study this in our previous paper). This can be explained using the approach we proposed previously\cite{qiaoCaptureRodlikeMolecules2020}: the orientational radius $R_\theta$ can be estimated by comparing the times needed by the rod to rotate due to diffusion and due to electrostatic forces. The rod's free rotational relaxation time is roughly the time it needs to diffuse over its own length, and thus scales like $\tau_\theta \sim L^2/D$. The force driving rotation at distance $r$ is $F(r) \sim \mathrm{d}\psi_\theta/L\mathrm{d}\theta\sim L\lambda_e k_BT/r^3$, where $\psi_\theta$ is the rotational potential energy\cite{qiaoCaptureRodlikeMolecules2020}; the corresponding time scale is $\tau_{\varepsilon} \sim L/(F/\gamma)$, where $\gamma = k_BT/D$ is the friction coefficient. The location $r$ where $\tau_\theta=\tau_\varepsilon(r)$ thus scales like $r \sim {(\lambda_eL^2)^{1/3}} \sim R_\theta$, irrespective of the presence of EHI.

\section{Conclusion and discussion}
We have built computational raspberry-like rods with different charge distributions and simulated their electrophoretic and mechanical drift in free solution under various salt conditions to investigate how electrohydrodynamic interactions impact the electrophoretic velocity and orientation of short rod-like charged molecules. We also studied how these interactions and the details of the charge distribution affect the capture, translocation, and escape of these rods. For instance, we tested our previously defined orientation radius $R_\theta$ in the new simulation setup, and studied the effect of the rod-wall interactions on rod capture. In order to be systematic, we present our main conclusions as a list of points below: 
\begin{itemize}
    \item The decrease of the free-solution electrophoretic mobility $\mu_e$ with salt concentration is a result of the competition between charge screening and rod orientation (Figs.~\ref{Fig:mu_C}a and c). Rods charged at one end only tend to orient with the field and have a higher mobility due to the lower frictional drag they encounter when oriented that way. For rods with other charge distributions,  increasing the salt concentration also orients the rod with the field and thus increases $\mu_e$. On the other hand, $\mu_e$ is reduced due to the charge screening at high salt concentration.
    
    \item The mechanical drift mobility $\mu_m$ is independent of the salt concentration (Fig.~\ref{Fig:mu_C}b) and is higher when the force is applied at the end of the rod. Rods with symmetric charge distributions move perpendicular to the force for the concentration range we have tested (Fig.~\ref{Fig:mu_C}c), which is consistent with what is reported in refs\cite{shinStructureDynamicsDilute2009,hamidNumericalStudySedimentation2020}.

    \item The rod velocity and the applied field/force are not necessarily pointing in the same direction even in a uniform field due to hydrodynamic interactions (Fig.~\ref{Fig:correlation_v_f}). The maximum deviation is $\approx 10^{\circ}$ for $C_s/C_{s}^o=0.03$ when the rod makes an angle of $\theta \approx \pi/3$ with the field.
    
    \item The rod's orientational order parameter is asymmetric during the capture and escape processes despite the field lines being identical on both sides of the wall. The charge distribution has more effect on rod orientation during the capture process when compared to the escape process. As we illustrated in a previous paper\cite{qiaoCaptureRodlikeMolecules2020}, the translational motion is too fast for a rod to rotate to its equilibrium orientation when it is within the orientational capture radius $R_\theta$. Similarly, the drift is too fast for the rod to lose its orientation due to thermal motion during the early phase of the escape process. Together, these two phenomena lead to asymmetric order parameter trajectories.

    \item However, the charge distribution impacts capture, translocation, and escape times, as well as the way rods enter the nanopore (Table~\ref{tab:2}). End-charged rods enter the nanopore via their charged end while there is no preference for other rods due to their symmetric charge distributions. The translocation time has a strong dependence on the charge distribution. For instance, sometimes the part of the rod that is inside the nanopore is uncharged, which severely slows down the translocation; for example, the translocation time is approximately 2.2 times larger for end-charged rods compared to double-helix rods (Table~\ref{tab:2}). Rods with charges only on their two ends orient faster during the capture process due to higher torque; however, they also lose orientation more rapidly after leaving the nanopore because the torque then amplifies the thermal fluctuations that make the head of the rod move away from the nanopore axis.  
    
    \item The capture time is correlated with the initial angular (polar) position and orientation of the rod when it starts close to the nanopore, in agreement with our previous investigations\cite{qiaoCaptureRodlikeMolecules2020}. The mean capture time of a rod is shorter when it is launched near the wall because the rod then tends to align along the wall\cite{paddingTranslationalRotationalFriction2010a,waszkiewiczHydrodynamicEffectsCapture2021}.
    
    \item The previously defined orientational capture radius $R_\Theta$ is still valid for both the capture and escape processes when EHI are present.

\end{itemize}

Overall, our simulations of the raspberry-like rod provide us with a more complete picture of the electrophoresis of rod-like molecules both in free solution and during translocation. We have demonstrated the important role that salt plays for rod orientation and charge screening, especially when the charge distribution is not uniform. Our results of rod orientation during capture qualitatively agree with our previous theories and LD simulations \cite{qiaoCaptureRodlikeMolecules2020} as well as with theoretical calculations\cite{waszkiewiczHydrodynamicEffectsCapture2021} that account for the anisotropic friction coefficient of rods and near-wall interactions. For applications such as the translocation of aptamer-bound molecules\cite{szeSingleMoleculeMultiplexed2017,reynaudSensingNanoporesAptamers2020}, our simulation results shine some light on the underlying physics under different conditions. On a different note, our results also suggest that one could enhance the capture rate of long flexible polymers by deliberately elongating their conformation; this might be achievable by attaching a slower, uncharged component to one end of the polymers, an idea that we are currently testing.


\section*{Acknowledgements}
GWS acknowledges the support of both the University of Ottawa and the Natural Sciences and Engineering Research Council of Canada (NSERC), funding reference number RGPIN/046434-2013. LQ is supported by the Chinese Scholarship Council and the University of Ottawa. The authors would like to thank Christian Holm and Kai Szuttor for their help with setting up the LB simulations and for fruitful discussions.




\bibliography{Refs} 

\end{document}